\documentstyle[12pt,epsfig,psfrag,amsmath,pifont,amssymb,cite]{article}

\numberwithin{equation}{section}
\numberwithin{table}{section}
\def\ga{\mathrel{\raise.3ex\hbox{$>$\kern-.75em\lower1ex\hbox{$\sim$}}}}
\def\la{\mathrel{\raise.3ex\hbox{$<$\kern-.75em\lower1ex\hbox{$\sim$}}}}

\def\tr{{\rm tr}}

\def\I_M{{I_{\scriptscriptstyle M\times M}}}

\def\be{\begin{equation}}
\def\ee{\end{equation}}
\def\bea{\begin{eqnarray}}
\def\eea{\end{eqnarray}}
\def\pa{\partial}

\newcommand{\beqal}{\begin{eqnarray}\label}
\newcommand{\beqa}{\begin{eqnarray}}
\newcommand{\eeqa}{\end{eqnarray}}
\newcommand {\expt}[1]{\left\langle #1 \right\rangle}

\newcommand {\f}{\frac}

\begin{document}

\begin{titlepage}
\begin{center}
\vskip .2in

{\Large \bf A note on matrix model with IR cutoff and AdS/CFT}
\vskip .5in

{\bf Tanay K. Dey$^a$\footnote{e-mail: tanay.dey@wits.ac.za}, 
Sudipta
Mukherji$^b$\footnote{e-mail: mukherji@iopb.res.in},
Subir Mukhopadhyay$^c$\footnote{e-mail: subirkm@gmail.com} and\\
Swarnendu Sarkar$^d$\footnote{e-mail: swarnendu@theory.tifr.res.in}\\
\vskip .1in
{\em $^a$National Institute for Theoretical Physics,\\
Department of Physics and Centre for Theoretical Physics,\\
University of the Witwatersrand,\\
Wits, 2050, South Africa.}
\vskip .1 in
{\em $^b$Institute of Physics,\\
Bhubaneswar 751~005, India.}
\vskip .1in
{\em $^c$Saha Institute of Nuclear Physics,\\
1/AF Bidhannagar, Kolkata-700~064, India}
\vskip .1in
{\em $^d$Department of Theoretical Physics\\ 
Tata Institute of Fundamental Research,\\ 
Mumbai 400~005, India.}}
\end{center}

\begin{center} {\bf ABSTRACT}
\end{center}
\begin{quotation}\noindent
\baselineskip 15pt
We propose an effective model of strongly coupled gauge theory at finite temperature on 
$R^3$ in the presence of an infrared cutoff. It is constructed by considering the 
theory on $S^3$ with an infrared cutoff and then taking the size of the $S^3$ to infinity 
while keeping the cutoff fixed. This model reproduces various qualitative features expected 
from its gravity dual.

\end{quotation}
\end{titlepage}
\vfill
\eject

\setcounter{footnote}{0}
\section{Introduction}

Considerable progress has been made recently in understanding various aspects of 
gauge theories with 
confinement \cite{witten,alvarez,rcharge,bobby,dmms1,dmms2,bst}. It turned out that one way 
of 
fabricating gravity dual of a gauge theory with confinement is to introduce an ultraviolet cut-off. Indeed, in \cite{ks}, the 
gravity dual of the confining gauge theory on flat space was constructed by smoothly capping 
off the geometry at small radius. A simpler model appeared in \cite{Ghoroku:2006cc,herzog} 
where the gravitational 
description consists of AdS$_5$ 
with small radius region removed and 
this was termed as a hard wall approximation. It is shown \cite{herzog} that for small enough cutoff (infact, smaller than the size of a black hole horizon), one obtains Hawking-Page 
transition at a finite nonzero temperature. 
Introduction the cut-off in a milder way leads to a soft wall approximation where one uses
an appropriate dilaton field in the background to truncate small radius region. 
However, in either cases, the short distance cutoff on the gravity side translates into an 
IR cutoff in the
corresponding gauge theory and so the gauge theory dual to the gravity theory under 
consideration is a theory on $R^3$
in the presence of an IR cutoff. 
Using the gravity dual one 
can learn various aspects of this gauge theory in the  strongly coupled regime. At this 
point one may ask: Can these lessons be used to design an  effective theory for the 
strongly 
coupled regime of this gauge theory with IR cut-off? This is essentially the point that we 
would like to address in the present note.

Necessary hint comes from the consideration of the  confining gauge theory on a compact 
space, such as, $S^3$. There one does not need a cut-off and radius of the three-sphere 
provides the scale essential for confinement.
Indeed an effective model for strongly coupled regime of this gauge 
theory on $S^3$ was constructed in  \cite{min,alvarez}. 
This model, which we call $(a,b)$ model in the sequel, is characterized by two parameters $a$ and $b$.
These parameters, in general, depend on temperature and the 't Hooft coupling. If $a, b$
satisfy certain properties, this model reproduces the qualitative features of the phase structures
expected from the dual theory in the bulk. To construct an analogus model on $R^3$ with an IR cutoff, 
we use the following route:

$\bullet$ ~ Incorporate the IR cut-off in the $(a,b)$ model so as to describe a gauge theory on $S^3$ with an infrared cutoff. 

$\bullet$ ~Keep the cutoff fixed and let the size of $S^3$ go to infinity.

The first step of constructing an intermediate model is easier. On generic ground, we 
expect that, in the presence of cutoff, the parameters $a$ and $b$ 
would depend on the cutoff. Indeed, we find that the $(a,b)$ model reproduces the  
qualitative features 
of gauge theory on $S^3$ with cutoff if $a, b$ depend on the cutoff in a specific way. 
This can be computed by comparing the effective potential 
in the gravity side with the free energy of the gauge theory arising from the $(a,b)$ model. 
In the absence of cutoff, $a$ and $b$ are monotonically increasing function of the temperature 
\cite{alvarez}. However, we find, while $b$ remains an increasing fuction even in the presence
of cutoff, the nature of the variation of $a$ is sensitive to the magnitude of the cutoff.
The second step is more difficult. Since the radius of the three-sphere does not appear as 
a tunable parameter, there is no simple procedure to carry out this step. Comparing gravity 
duals of the gauge theories on $S^3$ and $R^3$ respecvtively we propose that the effective model 
remains qualitatively similar. Based on this assumption we have constructed an effective 
model. Details of the computation is presented in latter sections of the note. Indeed our 
model posseses all the qualitative features expected from the gravity side. However, it 
also leads to a puzzle which we address in the next paragraph.

One of the features of our effective model is that there always exists an unstable maximum
which separates the two minima. These two minima, in the gravity side, represent thermal AdS and black 
hole in presence  of UV cutoff. On the other hand, one expects the maximum to represent a small 
unstable black 
hole. However, for the cases analysed in \cite{herzog}, such a black hole is missing. 
We suspect that this is due to the fact that the cutoff was introduced by hand. Once this cutoff is 
intruduced, both in soft and hard wall models dynamically, such black hole would prehaps 
appear. A preliminary support of our proposal comes from a calculation
carried out in the apendix of this paper. Here we analyse a gravity action which has
a fine tuned dilaton potential. If we set the dilaton to zero, we have a flat AdS black hole.
The temperature of this black hole increases linearly with the horizon size. However, as we turn
on a dilaton profile and the dilaton potential ( satisfying the BF bound), we see that for a given 
temperature, there can have two black holes -- one of which is unstable.

The plan of the paper is as follows. In section (\ref{grcutoff}), 
we analyze the gravity theory with a radial cutoff. In section 
(\ref{effzero}), we consider the zero coupling limit of gauge theory with an IR cutoff. 
In section (\ref{effstrongsphere}), we study the behaviour of the phenomenological 
matrix model on $S^3$ with an IR cutoff. Next section, (\ref{effstrongplane}), consists
of analysis of the model as we take the size of $S^3$ to infinity, keeping the cutoff
fixed. This paper ends with a discussion of our results. Finally, in the appendix,
we show, with in the soft wall model, how an unstable black hole may appear if we
introduce cutoff dynamically. This is a prediction that arise from an analysis of our
martix model of $R^3$. 

\section{Bulk with UV cutoff}\label{grcutoff}

We begin this section, with gravitational configuration having boundary $S^3\times S^1$ and its  thermodynamic properties 
the presence of an UV cutoff. In the latter part of this section, we take a suitable limit to obtain  gravitational configuration having boundary $R^3\times S^1$ and study its analogous properties.

$\bullet$ Gravitational configuration with boundary $S^3\times S^1$

Five dimensional gravity action with a negative cosmological constant is given by,
\begin{equation}
S = \frac{1}{2 \kappa^2} \int d^5 x {\sqrt{-g}} \Big( R + {\frac{12}{l^2}}\Big) ,
\end{equation}
where $\kappa^2$ is the five dimensional gravitational constant and $l^2$ parametrizes the 
cosmological constant. This action admits two solutions with same asymptotic behaviour: AdS 
and AdS black hole. For both, the metric can be written as
\begin{equation}
dS^2 = - V(r) dt^2 + {V(r)}^{-1} dr^2 + r^2 d \Omega_3^2.
\label{met}
\end{equation}
While for AdS, $V(r)$ is
\begin{equation}
V(r) = 1 + \frac{r^2}{l^2},
\label{vads}
\end{equation}
for the AdS black hole it has the form
\begin{equation}
V(r) = 1 + \frac{r^2}{l^2} - \frac{m}{r^2}.
\label{vhole}
\end{equation}
In the last equation $m$ is associated to the ADM mass of the black hole
and is given by
\begin{equation}
M = \frac{3 \omega_3 m}{\kappa^2},
\end{equation}
where $\omega_3$ is the volume of the unit three-sphere. In the asymptotic limit, $r \rightarrow \infty$, the metric becomes the AdS metric. The black hole has a curvature 
singularity at $r = 0$ and has a single horizon. The horizon radius 
$r_+$ can be obtained by setting $V(r)=0$ and is given by the real positive root of
\begin{equation}
1 - \frac{m}{r^2} + \frac{r^2}{l^2} = 0. \label{horizonradius}
\end{equation}

Various thermodynamic quantities for the black hole with an UV cutoff is given in \cite{cai}. 
Continuing to 
Euclidean signature, with time direction as a circle, one finds that smoothness of the 
geometry at non-zero $r$ requires the 
radius of the time circle to vanish at horizon. This leads to a temperature,
\begin{gather}
T = \frac{2 r_+^2 + l^2}{2 \pi r_+ l^2},
\label{temperature}
\end{gather} 
which is independent of the cutoff. 
There is a critical temperature $T_0$, termed as nucleation temperature, 
below which there is no black hole.  
For $T > T_0$ there are two black hole solutions - an 
unstable small black hole and a stable big black hole with horizon radius 
given by the smaller and larger root of $r_+$ respectively. They merge at $T=T_0$.

The free energy of the system is obtained by integrating the regularized action 
for AdS space and the black hole over the amputated space where a 
small three-sphere of radius $r_0$ is removed from the origin and 
so the lower limit of the integration on the radial coordinate is 
$r=r_0$ instead of the origin $r=0$. Imposing same asymptotic boundary 
condition for both the geometries and choosing the free energy of the AdS to be zero, 
one can obtain the free energy associated with the black hole as \cite{cai},
\begin{equation} 
F = -\frac{\omega_3}{2 \kappa^2 l^2} 
\Big( 2 r^4_{max} - r_+^4 - 2 r_0^4 - r_+^2 l^2\Big).
\label{freeenergy}
\end{equation}
Here we choose
\begin{gather}
r_{max} = r_+ \quad\quad{\rm for}\quad r_0 < r_+ \\ 
r_{max} = r_0 \quad\quad{\rm for}\quad r_0 > r_+.
\end{gather} 
Hawking-Page (HP) 
transition occurs only if $r_0 < r_+$ as only in this case the free energy 
(\ref{freeenergy}) changes sign.
In terms of Wilson loop operator, one can see that 
unless $r_0 < r_+$, there is no world-sheet belonging 
to zero homotopy group which ends on the temperature 
circle at the asymptotic infinity. Consequently, Wilson loop 
operator does not show a phase transition. The same conclusion is reported in \cite{herzog}. At this point it may appear in mind that since cut-off is within the horizon thermodynamic quantities are not affected by it. Indeed we will see the critical exponents do not depend on the cut-off. 

Since we are interested in comparing the thermodynamic behaviour with 
that of the matrix model counterpart 
coming from gauge theory, where the order parameter 
is the moment of the vev of the Wilson loop operator,  we restrict ourselves to $r_0 < r_+$. 
The situation for $r_0 \geq r_+$ is 
considered in\cite{cai}. The entropy of this sytem turns out to be
\begin{equation}
S = \frac{2 \pi \omega_3 r_+^3}{\kappa^2},
\label{entropy}
\end{equation}
and does not depend on the cutoff.

In order to summarize different phases in a 
convenient way ( which will help us in later sections to make a comparison with the matrix 
model ), we construct a Landau function. This reproduces equations
(\ref{temperature}, \ref{freeenergy}, \ref{entropy}) on shell.
First, we define the following dimensionless quantitites
\begin{equation}
\frac{r_+}{l} = \bar r, \frac{r_0}{l} = \bar r_0, {T l = \bar T}, 
\frac{\kappa^2}{ l^3} = \bar \kappa^2.
\end{equation}
Then, by identifying the horizon radius $\bar r$ as the order parameter,
it is easy to write down the Landau function ${\cal {F}}(\bar r, \bar r_0, \bar T)$ as, 
\begin{equation}
\bar {\cal {F}}(\bar r, \bar r_0, \bar T) = \frac{\omega_3}{2 \bar \kappa^2}
\Big(3 \bar r^4 - 4 \pi \bar T \bar r^3 + 3 \bar r^2 + 2 \bar r_0^4\big),
\label{lfun}
\end{equation}
where $ \bar {\cal {F}}(\bar r, \bar r_0, \bar T) = l {\cal {F}}(\bar r, \bar r_0, \bar 
T)$. On shell, this function reduces to the expression of free energy given in 
(\ref{freeenergy}). In order to see this, we need to compute the on shell temperature, 
which can be obtained by setting first variation of Landau function with respect to 
the order parameter $\bar r$ equal to zero. The result matches with (\ref{temperature}). 
Then substituting it in (\ref{lfun}) we get (\ref{freeenergy}). 
We plot the Landau function for various temperatures with 
three different values of cutoff. The critical temperature 
corresponds to having two 
degenerate minima of ${\cal{F}}$. This can be determined as follows. 
The function  ${\cal{F}}$ vanishes when the right hand side in equation (\ref{lfun})
is identically equal to zero. This, in turn, gives an equation for $T$ involving 
$\bar r$ and $\bar r_0$. Equating this with (\ref{temperature}), we get $\bar r$ in terms of
$\bar r_0$. Upon substituting this $\bar r$ in (\ref{temperature}), we finally get 
the critical temperature as
\begin{equation}
\bar T_c = \frac{2 + \sqrt {1 + 8 \bar r_0^4}}{ {\sqrt 2} \pi \sqrt{1 + \sqrt{1 + 8 \bar 
r_0^4}} }.
\end{equation}
%
\begin{figure}[ht]
\centerline{
\includegraphics[width=5.7cm]{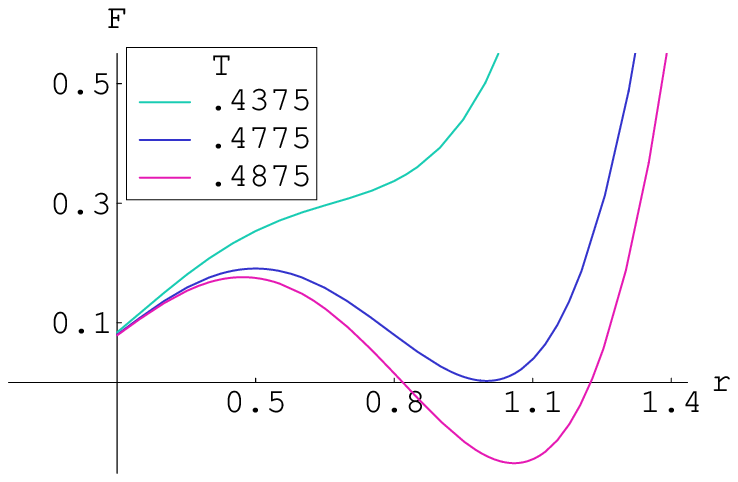}%
\hspace{-.02cm}
\includegraphics[width=5.7cm]{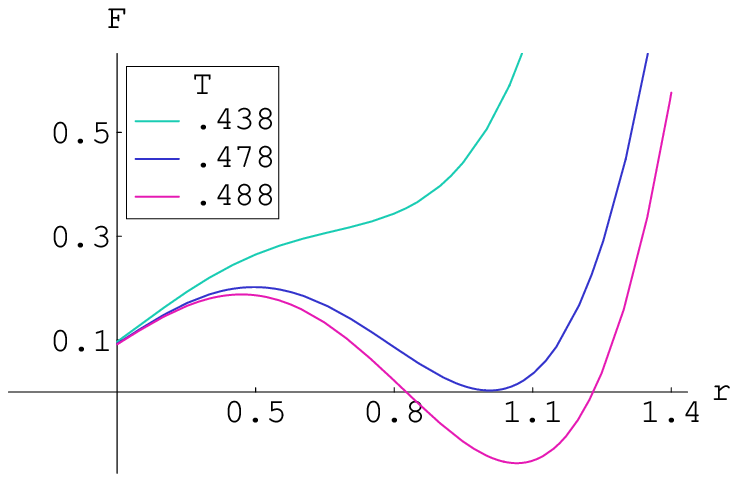}%
\hspace{-.02cm}
\includegraphics[width=5.7cm]{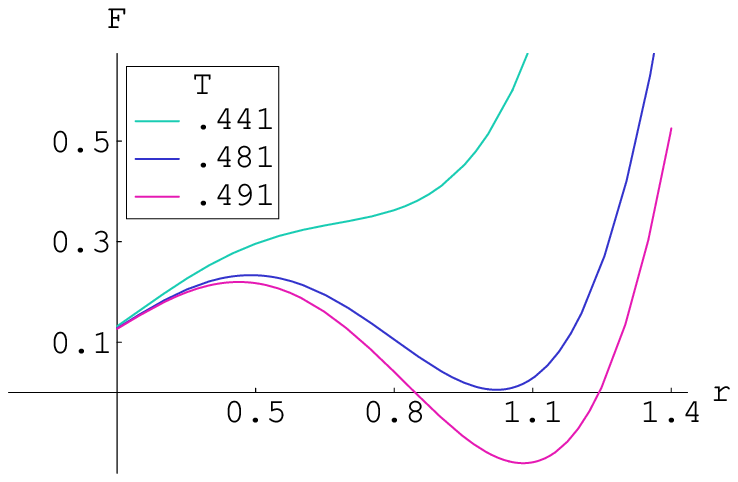}}
\caption{{\small{The plot of the Landau function ${\cal{F}}$ with
$\bar r$ for various temperatures with $r_0=0.2,\,0.3$ and $0.4$ from left to right.}}}
\end{figure}

From the plots we observe that the transition temperature increases 
with the cutoff. We will see that a similar behaviour occurs 
on the gauge theory side as well. Nevertheless, the qualitative 
features remain independent of the cutoff. 
In addition, we find that the  critical behaviour around the nucleation temperature ($T_0$), 
is universal and does not depend on details of the model. A computation of the specific heat 
near the nucleation temperature, $T_0$ shows that it
diverges as $(T-T_0)^{-1/2}$. The critical exponent is thus unaffected by the presence of 
the cutoff.

$\bullet$ Gravitational configuration with boundary $R^3\times S^1$

Having discussed the gravitational configurations with boundary $S^3 \times S^1$ in the
presence of an ultraviolate cutoff, we now review the case where the boundary is
$R^3 \times S^1$. This can simply be obtained by appropriate scaling of the coordinates
and the parameters that are involved in the previous geometries. We scale
\begin{equation}
r \rightarrow \lambda r, ~~ t \rightarrow \frac{l^2}{\lambda}t, ~~r_0 \rightarrow \lambda 
r_0, ~~ m \rightarrow \frac{\lambda^4 r_h^4}{l^2}.
\end{equation}
Then, taking the limit $(\lambda/l)\rightarrow \infty$, earlier black hole
metric reduces to
\begin{equation}
ds^2 = l^2 \Big[  \Big( r^2 - \frac{r_h^4}{r^2}\Big) dt^2 + \Big(r^2 -
\frac{r_h^4}{r^2}\Big)^{-1} dr^2 + r^2 \sum\limits_{i=1}^3 dx_i^2\Big].
\end{equation}
where $x_i$ are the appropriate coordinates on $R^3$.  The horizon radius in this limit 
reduces to $r_h$.

\begin{figure}[t]
\begin{center}
\begin{psfrags}
\psfrag{a}[][]{$\bar r$}
\psfrag{b}[][]{${\cal \bar F}$}
\epsfig{file=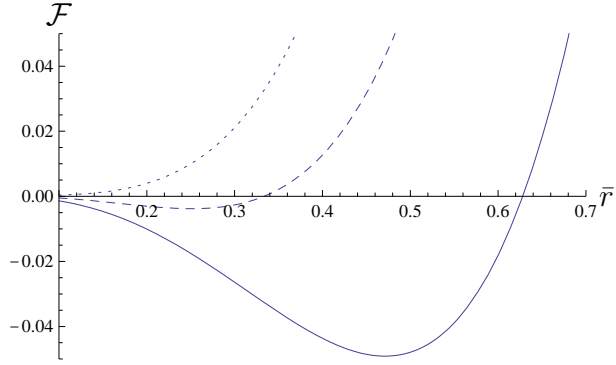, width= 8cm,angle=0}
\end{psfrags}
\vspace{ .1 in }
\caption{The plot of the Landau function $\bar{\cal{F}}$ with $\bar r$ for various temperatures with 
$r_0=0.1$.}
\label{landauflat}
\end{center}
\end{figure}


Since the temperature is related to the inverse of the Euclidean time, in this
scaling limit, the temperature of the black hole reduces to
\begin{equation}
T = \frac{r_h}{\pi}.
\end{equation}
We note that, unlike the case of black hole with spherical horizon, here for a
given temperature, there is only a single black hole with a non-zero horizon. The free
energy of this flat black hole follows from the spherical ones in the scaling limit
and is given by
\begin{equation}
F = -\frac{1}{2 \bar\kappa^2 l^2}(2 r_{\rm max}^4 - r_h^4 - 2 r_0^4),
\end{equation}
where
\begin{eqnarray}
 r_{\rm{max}} &=& r_h , \quad {\rm{for}} ~r_0 < r_h
,\nonumber \\
&=& r_0 , \quad {\rm {for}} ~r_0 > r_h ,\nonumber
\end{eqnarray}
and we have scaled the free energy $\lambda^4 F \rightarrow \tilde{F} $. The
Hawking-Page transition when $r_0 < r_h$ and at temperature $T_c = 2^\frac{1}{4} (r_0/\pi)$. 

Finally, the Landau function can be obtained easily as 
\begin{equation}
\bar{\cal{F}} = \frac{1}{{2 \bar \kappa^2}} (3 \bar r^4 - 4 \pi \bar T \bar r^3 + 2 \bar 
r_0^4),
\end{equation}
where we have defined $\bar r = r_h/l$. For $r_0=0.1$ the plot of this function for 
different temperature is shown in 
figure (\ref{landauflat}).

We end this section with the computation of the potential  for a quark-antiquark pair on 
$R^3\times S^1$. 
The confined phase of the gauge theory corresponds to a thermal AdS geometry in the bulk. We 
thus compute the Wilson loop  
 by computing the area of the dual world sheet extending into the bulk AdS space following 
\cite{Maldacena:1998im,Rey}. 
The computation will however be done in the pesence of the cutoff\footnote{Similar 
computations, with the bulk AdS Schwarzschild black hole geometry, were carried out in 
\cite{Andreev:2006eh,bbf} where phase transition different from the one studied in the 
present paper, was considered.}.

The AdS metric with boundary $R^3\times S^1$ is,

\beqal{adsmetric}
ds^2&=&\alpha^{'}\left\{\f{U^2}{R^2}\left(-dt^2+dx_i^2\right)+R^2\f{dU^2}{U^2}
+R^2 d\Omega_5^2\right\},\\
R^2&=&\sqrt{4\pi gN}.
\eeqa
The AdS radius is $l^2=\alpha^{'}R^2$.\\

To compute the potential between a quark and antiquark pair placed at $x=\pm L/2$, we consider
a rectangular Wilson loop in the $x-t$ plane. The dual area can be computed from the Nambu-Goto action,

\beqal{ng}
S=\f{1}{2\pi\alpha^{'}}\int d\sigma d\tau 
\sqrt{\det\left(G_{MN}\pa_{\sigma} X^M \pa_{\tau}X^N\right)},
\eeqa
where, $G_{MN}$ are the components of the metric (\ref{adsmetric}).
In the static gauge: $\sigma=x$ and $\tau=t$, we get

\beqal{ng1}
S=\f{T}{2\pi}\int dx \sqrt{(\pa_xU)^2+(U/R)^4}.
\eeqa

Equation of motion from (\ref{ng1}) satisfies,
\beqal{const}
\f{U^4}{\sqrt{(\pa_x U)^2 +(U/R)^4}}=C,
\eeqa
where $C$ is a constant that can be fixed by saying that $\pa_x U=0$ for $U=U_0$
($U_0$ is the cutoff radius). So that we get $C=R^2 U_0^2$.

\begin{figure}[t]
\begin{center}
\begin{psfrags}
\psfrag{A}[][]{A}
\psfrag{B}[][]{B}
\psfrag{C}[][]{C}
\psfrag{x}[][]{$x$}
\psfrag{x0}[][]{$x=0$}
\psfrag{U}[][]{$U$}
\psfrag{UC}[][]{$U_0$}
\psfrag{l}[][]{$\f{L}{2}$}
\psfrag{ml}[][]{$-\f{L}{2}$}
\epsfig{file=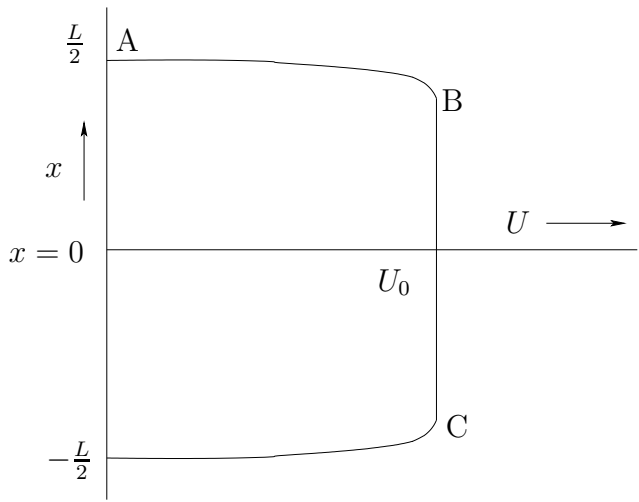, width= 8cm,angle=0}
\end{psfrags}
\vspace{ .1 in }
\caption{World-sheet configuration dual to the Wilson loop in the presence of cutoff, $U_0$.}
\label{wilson2}
\end{center}
\end{figure}

Equation (\ref{const}) can be solved for $x$ using the boundary condition that
$x (U=U_0)=x_0$,

\beqal{solx}
x-x_0= \f{R^2}{U_0}\int_1^{U/U_0} \f{dy}{y^2\sqrt{y^4-1}},
\eeqa
where $y=U/U_0$. For $U=\infty$, $x=L/2$. So given $L$ and $x_0$, one can solve for $U_0$ 
from (\ref{solx}).
The full energy of the configuration is given by the contribution from the sections
of the curve ``AB" plus ``BC", shown in figure (\ref{wilson2}).

The contribution from ``AB" is given by inserting the solution (\ref{solx}) into (\ref{ng1}),
\beqal{eab}
E_{AB}=\f{U_0}{2\pi}\left[\int_1^{\infty}\left(\f{y^2 }{\sqrt{y^4-1}}-1
\right)dy-1\right].
\eeqa
This contribution goes as the inverse power of $L$, because from (\ref{solx}),
$U_0 \sim R^2/(L/2 -x_0)$.
``BC" is the straight vertical line at $U=U_0$, the contribution from which is given by evaluating 
(\ref{ng1}) along this line,
\beqal{BC}
E_{BC}=\f{1}{4\pi}\left[L -L_0\right].\f{U_0^2}{R^2},
\eeqa
where, in equation (\ref{BC}), $L_0/2= L/2-x_0$. The string tension is thus,
\beqa
\sigma=\f{1}{4\pi}\f{U_0^2}{R^2}.
\eeqa
Going to the original radial coordinate by replacing $U=r/\alpha^{'}$,
\beqa
\sigma=\f{1}{4\pi\alpha^{'}}\f{r_0^2}{l^2}\label{strt}.
\eeqa
This shows that the tension grows with the cutoff. For ${\cal N}=4$ SYM
on $S^3$ we do not expect to get such a linear potential and hence a 
string tension. This potential has been recently computed in \cite{BallonBayona:2008uc}. 
Though the nature of phase transition in the two cases - that is one on 
finite $S^3$ and the other on $R^3$ with a cutoff are similar, the behavior 
of the potential before and after the transitions seem to be quite different in the two 
cases. 
  
Let us note that the string tension vanishes as the cutoff is taken to zero in equation 
(\ref{strt}).
This is expected because, in this case, the theory is same as that of one on the three-sphere
with the radius of the sphere taken to infinity where we only have the coulombic term
in the potential. Finally, if we consider the theory on $S^3$ with an additional cutoff, 
this $R^3$ result can also be thought of arising when $L/R_{S^3}<<1$, where $R_{S^3}$ is 
the radius of the three-sphere.

\section{Effective model at zero coupling with IR cutoff}\label{effzero}

In this section, we discuss an effective model obtained from gauge theory on a three-sphere in presence of an IR cut-off. A three-sphere provides a natural scale which translates into confinement scale and in view of that it may appear superfluous to consider an additional cut-off. However, as emphasized in the introduction this model is an intermediate step to construct an effective theory for strongly coupled gauge theory in $R^3$ in presence of an IR cut-off.
Before we move to the discussion of the boundary gauge theory at strong coupling, in this 
section 
we will begin with the partition function 
of ${\cal N}=4$ $U(N)$ super Yang-Mills theory at zero coupling on a three-sphere with an 
IR cutoff following \cite{min}.
The field content of the theory consists of vector bosons, scalar bosons and 
fermions all in adjoint representation of $U(N)$. We expand various fields 
into spherical harmonic modes on the three-sphere leading to a quantum mechanical 
system consisting of infinite number of degrees of freedom. The non-zero modes 
lead to massive degrees of freedom and it turns out that the zero modes are 
also massive in this problem. The fermionic zero modes are massive because 
of anti-periodic boundary condition around the temperature circle and the 
scalar zero modes are massive due to their coupling with the Ricci scalar 
of the three-sphere. Since we set the coupling to zero these degrees of freedom 
can be thought of as infinite number of decoupled harmonic oscillators 
each in the adjoint representation of $U(N)$. They can be either bosonic or 
fermionic depending on their parent field. Any excitation of this system, then, 
composed of single particle states corresponding to the various oscillator modes 
and a general physical state can be obtained by acting with an arbitrary collection 
of these oscillators on the Fock space vacuum. However, the Gauss's law does 
not allow any non-singlet state to be physical. So while considering 
physical excitations we need to consider those which are singlets of $U(N)$.

We begin with the partition function for bosonic modes in adjoint representation of $U(N)$. 
Single particle modes are denoted by $i$ with energy $E_i$. 
A generic state is represented by a set of integers $\{n_i\}$ where 
$n_i$ is the level of $i$'th mode and so with energy $\sum\limits_{i=0}^\infty n_i E_i$. 
The corresponding partition function can be written as
\begin{gather}
Z(x) = \prod\limits_i \sum\limits_{n_i=0}^\infty ~ x^{n_iE_i} S(n_i), \label{pfc1}
\end{gather} 
where $S(n_i)$ represent number of singlet states in a symmetric product of $n_i$ number of particles in adjoint representation, which takes care of the fact that we are summing over physical states only.
Since number of singlet in any representation can be obtained by integrating the character of that representation over the group manifold, we write
\begin{gather}
S(n) = \int [dU] \chi_{\text{sym}^n(\text{adj})}.
\end{gather}
We substitute this integral expression in (\ref{pfc1})
to obtain
\begin{gather}
Z(x) = \int [dU] \prod\limits_i \sum\limits_{n_i=0}^\infty ~ x^{n_iE_i} \chi_{\text{sym}^{n_i} (\text{adj})}.
\end{gather}
As shown in the appendix of \cite{min}, this expression reduces to 
\begin{gather}
Z(x) = \int [dU] \exp \left[ \sum\limits_i \sum\limits_{m=1}^\infty \frac{1}{m} x^{mE_i} tr(U^m)tr((U^\dagger)^m) \right], \label{pfc2}
\end{gather}
where we use
$\chi_{\text{adj}} = tr(U)tr(U^\dagger)$ and the $tr$ implies trace in the fundamental representation.
We can carry out a partial sum over all single particle states associated with a particular field. There could be different single particle partition functions for scalars and vectors. For the time being we assume only one kind and write single particle partition function as $z(x) = \sum\limits_{i} x^{E_i}$. The above partition function in (\ref{pfc2})reduces to
\begin{gather}
Z(x) = \int [dU] \exp \left[ \sum\limits_{m=1}^\infty \frac{1}{m} z_B(x^m) tr(U^m)tr((U^\dagger)^m) \right].
\end{gather}
A similar computation can be done for the fermionic fields also (see\cite{min} for details) where the symmetric product in (\ref{pfc1}) is replaced by antisymmetric product. 
Considering both the bosons and the fermions we obtain the total partition function as
\begin{gather}
Z(x) = \int [dU] \exp \left[\sum\limits_{m=1}^\infty \frac{1}{m} [ z_B(x^m) + (-1)^{m+1} z_F(x^m)] tr(U^m)tr((U^\dagger)^m) \right] ,
\label{pfc3}\end{gather}
where $z_B$ and $z_F$ are bosonic and fermionic single particle partition function respectively. If we have $n_B$ number of bosonic fields and $n_F$ number of fermionic field the expression in the square bracket in (\ref{pfc3}) will be replaced by $n_B z_B(x^m) + (-1)^{m+1} n_F z_F(x^m)$. In addition, if there are two kinds of bosonic fields, such as, $n_v$ number of vector fields and $n_s$ number of scalar fields the term becomes
$n_v z_v(x^m;R) + n_s z_s(x^m;R) + (-1)^{m+1} n_F z_F(x^m)$ where $z_v$ and $z_s$ are single particle partition function for vector and scalar fields. 

In presence of a cutoff $\epsilon$ we will consider the states, 
with energy satisfying the bound $\sum\limits_{i=0}^\infty n_i E_i \geq \epsilon_0$ . The corresponding partition function can be written as
\begin{gather}
Z(x;\epsilon_0) = \prod\limits_i \sum\limits_{n_i=0}^\infty ~ x^{n_iE_i} \theta(n_i E_i - \epsilon_0) S(n_i),
\end{gather} 
where we use the $\theta$-function satisfying $\theta(x)=0$ for $x<0$ and $\theta(x)=1$ for $x>0$. Using the integral representation
\begin{gather}
\theta(x) = \int\limits_0^\infty \frac{dk}{k} e^{ikx},
\end{gather}
we can rewrite the partition function as 
\begin{gather}
Z(x;\epsilon_0) = \int \frac{dk}{k} e^{-ik\epsilon_0} \prod\limits_i \sum\limits_{n_i=0}^\infty ~ (xe^{ik})^{n_iE_i} S(n_i),
\\ = \int \frac{dk}{k} e^{-ik\epsilon_0}Z(xe^{ik}).
\end{gather} 
Then following the similar procedure as above we can write the partition function with cutoff from (\ref{pfc2}) by replacing $z_B$ and $z_F$ by $z_B^{\epsilon_0}$ and $z_F^{\epsilon_0}$ respectively
\begin{gather}
Z(x;\epsilon_0) = \int [dU] \exp \left[ \sum\limits_{m=1}^\infty \frac{1}{m} [ z_B^{\epsilon_0}(x^m) + (-1)^{m+1} z_F^{\epsilon_0}(x^m) ]tr(U^m)tr((U)^\dagger)^m) \right], \label{pfc5}
\end{gather}
where $z_B^{\epsilon_0}$ and $z_F^{\epsilon_0}$ are bosonic and fermionic single particle partition functions respectively with 
cutoff $\epsilon_0$. Both of them have the general form
\begin{gather}
z^{\epsilon_0}(x) = \int \frac{dk}{k} e^{-ik\epsilon_0}z(xe^{ik}). \label{pfc6}
\end{gather}

A formal path-integral derivation of the partition function for Yang-Mills theory at zero coupling is done 
in \cite{min} and turn out to give rise to the same partition function as in (\ref{pfc3}). All the modes 
once again gets massive and can be systematically integrated out. The only massless mode that remains in 
the spectrum is the zero mode associated with $A_0$ around the thermal cycle. Thus the group theoretic 
quantity $\frac{1}{N}Tr(U)$ actually represents Polyakov loop operator around thermal cycle and serves as 
an order parameter. 
In presence of cutoff the generalization of computation of partition function in a path-integral 
formulation is very similar and so we are not repeating that here. One will get the same result as (\ref{pfc3}).
 
The only variable in this effective action is an $N\times N$ unitary matrix and thus we have arrived at an effective matrix model.
Following \cite{min} we rewrite it in terms of the eigenvalues of $U$ as 
$\{e^{\alpha_i}\}$ where $\alpha_i$'s are within the range $(-\pi < \alpha_i \leq \pi)$ and obtain a system of eigenvalues distributed over a circle. 
The measure can be rewritten in terms of the eigenvalues as
\begin{gather}
\int [dU] \rightarrow \prod\limits_i \int\limits_{-\pi}^\pi [d\alpha_i]\prod\limits_{i<j} \sin^2(\frac{\alpha_i - \alpha_j}{2}).
\end{gather}
We make the substitution
\begin{gather}
\text{tr}(U^n) \rightarrow \sum\limits_j e^{in\alpha_j},
\end{gather}
in the action and 
write down the partition function as
\begin{gather}
Z(x;\epsilon_0) = \int [d\alpha_i] \exp[-\sum\limits_{i\neq j} V^\epsilon_0 (\alpha_i - \alpha_j)],
\end{gather}
where
\begin{gather}
V^{\epsilon_0}(\alpha) = -\log|\sin(\alpha/2)| - \sum\limits_{n=1}^\infty \frac{1}{n} [ z_B^{\epsilon_0} (x^n) + (-1)^{n+1} z_F^{\epsilon_0} (x^n) ] \cos(n\alpha).
\label{cfpot1}\end{gather}
The temperature independent repulsive potential is independent of 
cutoff and the attractive potential, which monotonically increases with temperature, decreases as the value of cutoff increases. 
That is what we expect because increasing the cutoff amounts to discarding lower powers of $x$ in a Taylor series expansion of $z_B(x)$ and $z_F(x)$ and so the attractive potential dominates over the repulsive potential at higher value of $x$.

This system of eigenvalues shows two different phases in terms of the eigenvalue distribution as order parameter. At low temperature the repulsive potential dominates and the eigenvalues tend to be distributed uniformly over the circle $-\pi \leq \alpha \leq \pi$. As temperature increases the attractive potential gets stronger and at some critical temperature the eigenvalues tend to clump together signalling a phase transition as happens in absence of cutoff. Since attractive potential decreases with cutoff while repulsive potential is independent of it, the phase transition temperature increases.
Introducing the eigenvalue distribution function, $\rho(\alpha)$, normalized as $\int\limits_{-\pi}^\pi d\alpha \rho(\alpha) = 1$ 
the effective action can be written as
\begin{gather}
S[\rho(\alpha)] = N^2 \int d\alpha \int d\beta \rho(\alpha) \rho(\beta) V^{\epsilon_0}(\alpha - \beta) = \frac{N^2}{2\pi} \sum\limits_{n=1}^\infty |\rho_n|^2 V_n^{\epsilon_0}, \label{cfeffaction}
\end{gather}
where $\rho_n := \int d\alpha \rho(\alpha)\cos(n\alpha)$ represents moments of $\rho$ and the moments of potential $V_n = \int d\alpha V(\alpha) \cos({n\alpha})$ are
\begin{gather}
V^{\epsilon_0}_n = \frac{2\pi}{n} ( 1 - z_B^{\epsilon_0}(x^n) - (-1)^{n+1} z_F^{\epsilon_0} (x^n) ).\label{vn}
\end{gather}
From (\ref{cfeffaction}) we see the uniform distribution is minimum when $V^{\epsilon_0}_n > 0$. As we increase the temperature it is $V_1$ which reaches the zero first and so the uniform distribution will be destabilized. Setting  $V^{\epsilon_0}_1(x_H) = 0$ we get at $x=x_H$ 
$z_B^{\epsilon_0}(x_H) + z_F^{\epsilon_0} (x_H) = 1$. This temperature is identified with the Hagedorn temperature of the corresponding gauge theory. The coefficients $z_B$ and $z_F$ depends on the bosonic and fermionic matter contents respectively. In particular,
for ${\cal N}=4$ SYM, in which we are interested, the single particle partition functions are given by
\begin{gather}
z_B(x) = \frac{6x + 12x^2-2x^3}{(1-x)^3},
\quad\quad
z_F(x) = \frac{16x^{3/2}}{(1-x)^3}.
\end{gather}
From this expression, using (\ref{pfc1}) we can calculate $z_B^{\epsilon_0}$ and $z_F^{\epsilon_0}$ respectively.
We have plotted the Hagedorn temperature for free ${\cal N}=4$ SYM against the cutoff and observe that 
the temperature decreases with increase in cutoff.
\begin{figure}[ht]
\epsfxsize=8cm
\centerline{\epsfbox{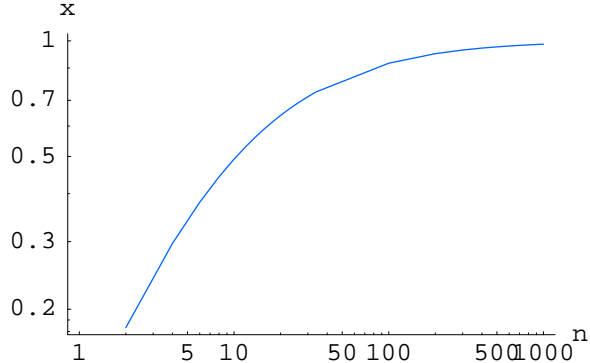}}
\caption{{\small{The plot of $x_H= e^{-(1/T_H)}$ with cutoff $\epsilon_0=n$. For convenience we have used a LogLog Plot.}}} 
\end{figure}

\section{Effective model at strong coupling on $S^3$ with IR}\label{effstrongsphere}

In the last section we briefly reviewed the derivation of the matrix model associated with an ${\cal N}=4$ SYM theory with an IR cutoff on a three-sphere at zero coupling. In order to compare with gravity  
we need to go to the strongly coupled regime. Let us begin 
considering the analogous analysis\cite{min,alvarez} in absence of cutoff. 
At weak coupling the Polyakov loop operator will remain the only massless 
mode and one can write down an effective action by integrating all other modes out. 
The effective action cannot be written in a closed form a priory, 
but from the requirement of gauge invariance and Gauss's law one can show that 
it will contain terms like $tr(U^{n_1})...tr(U^{n_k})tr(U^{-n_1-...-n_k})$. As evident from 
(\ref{cfeffaction}) at weak coupling mass of $n$-th  moment of distribution function $\rho$ is given by 
$V_n$. Since at a fixed temperature $V_1$ is the smallest, we expect that $\rho_1 =\frac{1}{N}tr(U)$ would 
be the lightest mode. 
So at weak coupling it is possible to integrate out all other moments and write down an effective action in 
terms of $\rho_1$ only. One extrapolates this 
to strong coupling 
and assumes so far the universal features of the system is concerned, 
one can write down an effective matrix 
model action quadratic in $\rho_1$ with coefficients  
depending  on the temperature the `t Hooft coupling $\lambda$. 

On the basis of the above arguments a matrix model consisting of 
a unitary matrix was introduced in \cite{alvarez}. The partition 
function and effective action are given by
\beqa
Z(\lambda,T)&=&\int dU e^{S_{eff}\left(U\right)},\\
S_{eff}(U)&=& a(\lambda,T) \tr \left(U\right) \tr \left(U\right)^{\dagger}+\frac {b(\lambda,T)}{N^2}
\left(\tr \left(U\right) \tr \left(U\right)^{\dagger}\right)^2,\label{effectiveaction}
\eeqa
respectively, which involves the order parameter characterising the deconfined phase of the gauge
theory given by the expectation value of the Polyakov loop
$(1/N) \expt{\tr U}$.
The equations of motion that follows from the action 
can be written as,
\begin{eqnarray}\label{saddle}
a\rho+2b\rho^3&=&\rho \mbox{\hspace{0.7in}} 0\le \rho \le \frac{1}{2},
\nonumber\\
&=&\frac{1}{4(1-\rho)} \mbox{\hspace{0.2in}} \frac{1}{2}\le \rho\le 1,
\end{eqnarray}
where $\rho^2=(1/N^2) \tr U \tr U^{\dagger}$. 
For later use let us note that the equations of motion can be integrated in the form of a potential. In particular, for $\rho > 1/2$ the potential can be written as
\begin{equation}\label{pot}
V(\rho) = 2a \rho^2 + 2b \rho^4 + \log(1-\rho) + f ,
\end{equation}
where we have added a constant, $f=\log(2)-1/2$ to make the potential
continuous at $\rho=1/2$.

Considering this model as an effective theory in the strong coupling regime,
the phases were studied in\cite{alvarez}. It was shown
that this model undergoes two phase transitions as a function of temperature.
There is a first and a third order transition when $b>0$ and $a<1$. 
 The thermal history matches with that of the bulk 
gravity as long as both $a$ and $b$ are increasing functions of 
temperature.
The first order transition can be mapped to the HP transition between
thermal AdS and the big black hole. It was further proposed that the third 
order transition corresponds to the black-hole string transition for the small
black hole. This transition is not visible in the gravity approximation of 
string theory.

Since on the gravity side, turning on the cutoff does not  
lead to any drastic change in the thermodynamic 
behaviour it is a reasonable assumption that the above  
arguments leading to the quadratic matrix model 
effective action will remain valid for non-zero cutoff  
and it is possible to describe the thermodynamic 
behaviour with the effective action given in (\ref{effectiveaction}),  
where the coefficients $a$ and $b$ 
will depend on the value of the cutoff as well. 
In what follows, we analyse this matrix model and comparing 
with the results obtained from gravity side we
study the behavior of
the coefficients $a$ and $b$ as functions of infrared cutoff.
For convenience we write the cutoff parameter as $\Lambda_{IR}$,
which is related to the cutoff  
$r_0$ on the gravity side through the relation 
$\Lambda_{IR} l^2 \sim r_0$. 

We now use a method similar to \cite{dmms1} in  order to study the temperature dependence of $a, b$
at various IR cutoffs. Subsequently, we discuss, with a mild assumption, how this method also allow us 
to find $\lambda$ dependence of $a$ and $b$ at fixed temperature. The procedure is as follows. 
On the matrix model side, at a generic temperature, there 
are two saddle points arising as the solution of (\ref{saddle}). 
Of which the unstable one (maximum) corresponds to the small black hole 
and the stable one (minimum) corresponds to the large black hole on the 
gravity side. We identify the matrix model
potential at maximum and minimum with the gravity action evaluated with the 
smaller and larger values of horizon radius respectively.
That will lead to the following two equations
\begin{gather}
2 a \rho_1^2 + 2 b \rho_1^4 + \log (1-\rho_1) + f = -I_1(r_0), \label{numeric1}\\
2 a \rho_2^2 + 2 b \rho_2^4 + \log (1-\rho_2) + f = -I_2(r_0),
\label{numeric2}
\end{gather}
where the $I_{1,2}(r_0)$ are the actions for the large and 
small black-holes
respectively and $\rho_{1,2}$ are the corresponding solutions in the
matrix model.  
Since we need the black hole solutions on the gravity side and 
the saddle points on the matrix model these equations are valid at 
$T > T_0$. At the black hole nucleation point, where $T=T_0$, above two equations converges into one. 
Therefore we can begin from an input value of $a(T_0)$ while 
$b(T_0)$ will be determined by that equation.
In addition to (\ref{numeric1},\ref{numeric2}) 
$\rho_1$ and $\rho_2$ satisfy the saddle point equation (\ref{saddle}) as well which leads to two more 
equations one for each of them.
\begin{gather}
a \rho_1 + 2 b \rho_1^3= \frac{1}{4(1-\rho_1)}, \label{numeric3}\\
a \rho_2 + 2 b \rho_2^3= \frac{1}{4(1-\rho_2)}. \label{numeric4}
\end{gather}
This set of four equations given by (\ref{numeric1},\ref{numeric2},\ref{numeric3},\ref{numeric4}) 
can be used numerically to eliminate
$\rho_1$, $\rho_2$, and to solve for $a$ and $b$. 
\footnote{We have set $l$, $\omega_3$, 
$\kappa_5$ to $1$ in the numerical computations.} and 
thus provide the temperature dependence of $a$ and $b$.

\begin{figure}[ht]
\begin{center}
\epsfig{file=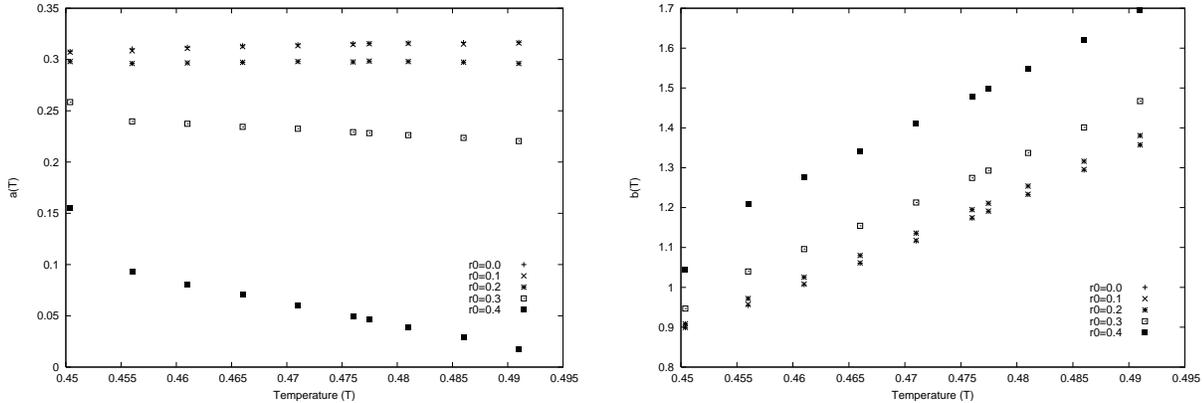, width= 16cm,angle=0}
\caption{{\small{Plot of $a(T)$ and $b(T)$ for various values of cutoff $r_0$}}}
\label{atbt}
\end{center}
\end{figure}

The variations of $a$ and $b$ as  function of temperature for various
values of $r_0$ are shown in figures (\ref{atbt}). In absence of cutoff both $a(T)$ and $b(T)$ are monotonically increasing function of temperature as shown in \cite{alvarez}. From figure (\ref{atbt}) we observe that $b(T)$ remains a monotonically increasing function of 
$T$ for all non-zero values of the cutoff $r_0$. However, from figure we find that $a(T)$ increases 
slowly for $r_0 < 0.1$. It becomes a decreasing function of 
temperature for higher values of the cutoff. This is somewhat puzzling
from the following point of view. At zero coupling, $a(T)$ has an 
interpretation of a single particle partition function. This increases 
with temperature. Thus one expects, even at a very weak coupling, $a(T)$ 
would remain a monotonically increasing function of 
temperature. In the absence of cutoff, we have also seen in \cite{dmms1} 
that the gravity data suggests $a(T)$ continues to increase with $T$ even 
in the strong coupling. In view of this one may have expected similar 
behaviour of $a(T)$ even in the presence of a cutoff. However, what we 
find here does not match with this expectation.


\begin{figure}[t]
\begin{center}
\begin{psfrags}
\psfrag{A}[][]{N}
\psfrag{B}[][]{D}
\psfrag{a}[][]{$$}
\psfrag{b}[][]{$$}
\epsfig{file=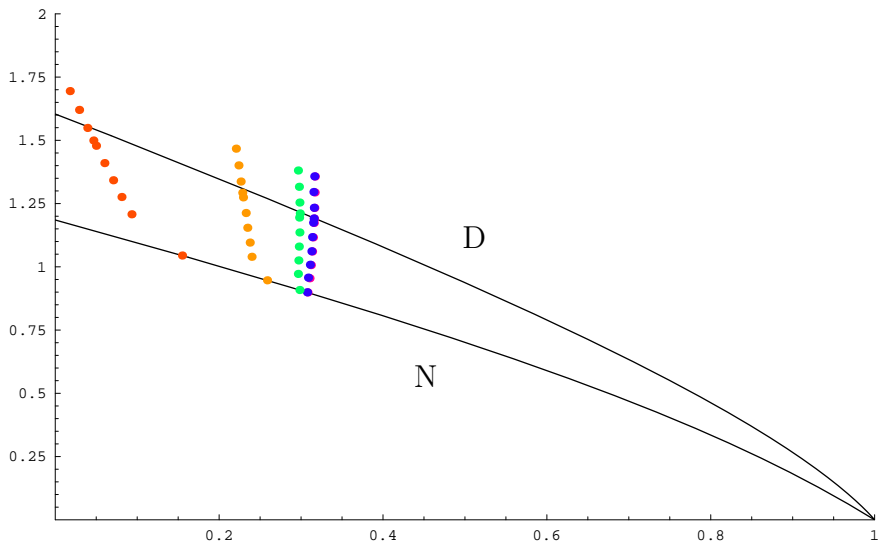, width= 12cm,angle=0}
\end{psfrags}
\vspace{ .1 in }
\caption{Thermal paths in the ($a-b$) space corresponding to the various values 
of the cutoff in gravity, ($r_0$) which increases from blue towards 
red. The direction of flow as the temperature 
increases is from bottom towards top. The vertical axis is $b$.}
\label{ab}
\end{center}
\end{figure}

In the figure (\ref{ab}), we plot the thermal paths in the ($a-b$) space with $b$ along the vertical-axis. 
Each dot in the figure has the coordinate $a(T)$ and $b(T)$ for a given value of cutoff $r_0$. the line ``N'' denotes the ``Nucleation'' point ($a(T_0)$, $b(T_0)$) of the blackholes in the ($a-b$) paremeter space. Similarly, line ``D'' is the``Deconfinement'' transition 
line corresponds to ($a(T_H)$, $b(T_H)$). For a given value of cutoff,
the dot moves upward with increasing 
temperature begining from the nucleation line. All dots on the nucleation line are at the same 
temperature independent of cutoff since the nucleation temperature of black holes in the bulk 
is unaffected by the cutoff while it is different on the 
deconfinement line. 

From the plot, following the dots along the path, one can see that deconfinement 
temperature increases with
increasing $r_0$. From the matrix model side, it can be seen that as cutoff increases 
the deconfinement temperature will increase  as long as $b$ remains 
an increasing function. We can compare this with the zero coupling situation.
There the transition temperature is determined by  
$a(T)=1$. As we raise the value of the cutoff, $a(T)$ increases 
more and more slowly with temperature and as a result the transition 
temperature increases. The free case can be thought of as a flow 
of $a(T)$ along the horizontal axis where $b=0$.  

The theory in this section has three scales, the temperature $\beta$, the radius of the 
sphere $R_{S^3}$ and the cutoff $R_c$ (which is related inversely to $r_0$ from the gauge/gravity duality).
$a$ and $b$ are thus functions of two dimensionless quantities: $\beta/R_c$ and $\beta/R_{S^3}$.
However it can be seen from Figure (\ref{atbt}) that the terms containing 
$R_{S^3}^2/(\beta R_c)$ dominate over the other dimensionless quantities. For fixed $R_{S^3}$,
both $a$ and $b$ scale with respect to $\beta$ in the same way as the cutoff $r_0$.
This dependence on two dimensionless combinations is in contrast to the theory without the cutoff 
with boundary $S^3\times S^1$ or the one with flat boundary and a cutoff. 
In the latter cases the free energy in gravity and 
hence the parameters $a$ and $b$ show much simpler dependence on the scales. 

The critical exponent around the nucleation temperature, $T=T_0$ for this matrix 
model has been calculated in \cite{alvarez}. This is obtained by expanding the
effective potential (\ref{pot}) in powers of $(T-T_0)/T_0$. As there is no 
explicit dependence of the parameters on the cutoff here, the critical exponents
thus remain the same. This is consistent with what we found in gravity, that the 
various critical exponents are insensitive to the cutoff.
 
We now use these above inputs to construct a model on $R^3$ with an infrared cutoff. This is the 
subject of the next section.


\begin{figure}[t]
\begin{center}
\epsfig{file=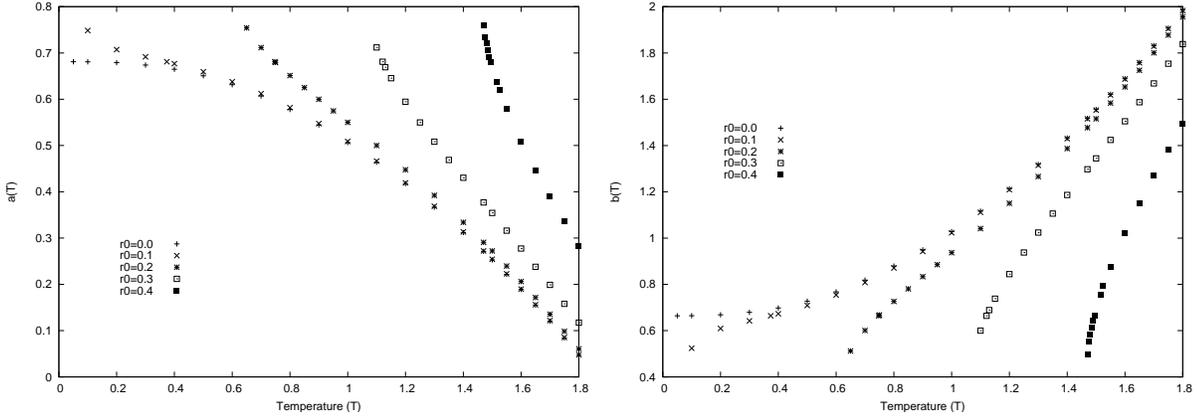, width= 16cm,angle=0}
\caption{{\small{Plots of $a(T)$ and $b(T)$ with various values of cutoff $r_0$ for the theory on $R^3\times S^1$.}}}
\label{atbtflat}
\end{center}
\end{figure}

\section{Effective model at strong coupling on $R^3$ with IR}\label{effstrongplane}

Having discussed the theory on $S^3$ with an additional IR cutoff, let us now
study its behavior as we take the radius of $S^3$ to infinity keeping the IR
cutoff fixed. The effective theory with two parameters $(a,b)$ serves as phenomenological  
model for the theory on $S^3 \times S^1$. The radius of the $S^3$ acts as an
infrared cutoff and there is a phase transition at a temperature proportional 
to the inverse of the radius. As the radius of the sphere is taken to infinity
the theory is always in the deconfined phase. We have seen in section (\ref{grcutoff}) that
for a flat boundary with an IR cutoff introduced by hand also shows the
same qualitative features as the theory on the sphere. Thus at the level
of a phenomenological model, it is natural to expect that
(\ref{effectiveaction}) should serve as good effective model 
for the theory on flat space with an IR cutoff. 
From the point of view of the boundary theory this relies on the assumption
that the effective model with IR cutoff also has an expansion in terms of the
Polyakov loop. The dimensionless quantity on which
the phases of the theory depend is $\beta/R_c$, where $R_c$ is
IR cutoff on the boundary. For the theory on the sphere $R_c=R_{S^3}$. 

The main qualitative difference as seen from section (\ref{grcutoff}) is that for the
flat boundary, the gravity theory with cutoff shows only one black hole solution that exists for all
temperatures, whereas the theory with the boundary $S^3$ has two. In the latter case, the euclidean 
version of this unstable black hole serves as the bounce for the AdS to black hole transition.
In $R^3$ with cutoff, one may expect a similar phenomenon when
we can find an exact solution of the Einstein's equation with an inbuilt cutoff 
(instead of putting it by hand). The Lorenzian version of the missing bounce would be 
an unstable black hole. In the appendix, we provide such a construction for the soft wall 
model.

From the matrix model side we have seen that it is impossible to write down a 
smooth potential that interpolates between two minima (corresponding to thermal AdS and the 
big black hole) without having an unstable maximum in between. Here in this analysis
we will include this unstable maximum so that it lies in the range $0\le \rho \le 1/2$.
The interpretation of such a saddle point is that of a stringy solution not visible in the
supergravity approximation. A transition to such a string phase from the black hole phase
was interpreted in \cite{alvarez,bst} as the gravity dual of the third order phase transition \cite{gw,w}
on the boundary. In a complete gravity solution with cutoff we expect to see an unstable
black hole as mentioned in the earlier paragraph the corresponding saddle point for which in the 
matrix model would lie in the region $\rho > 1/2$.

To implement this in the following numerical analysis
similar to that in section (\ref{effstrongsphere}), we keep the unstable saddle point at $\rho=0.49$. 
The 
following plots of $a$ and $b$ (Figure(\ref{atbtflat})) reflects the fact that they are functions of the
dimensionless quantity $\beta/R_c$.

\begin{figure}[t]
\begin{center}
\begin{psfrags}
\psfrag{A}[][]{N}
\psfrag{B}[][]{D}
\psfrag{a}[][]{$a$}
\psfrag{b}[][]{$b$}
\epsfig{file=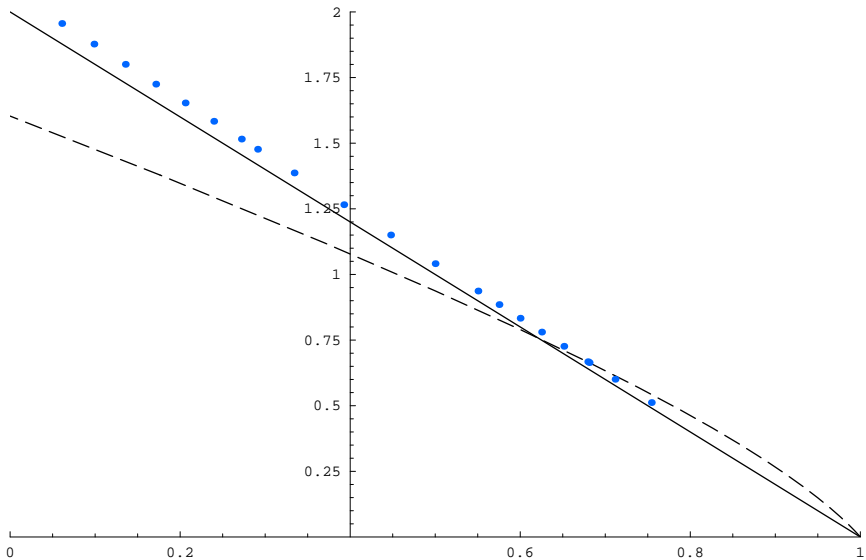, width= 12cm,angle=0}
\end{psfrags}
\vspace{ .1 in }
\caption{The thermal path in the ($a-b$) space for all values 
of the cutoff in gravity, ($r_0$). The direction of flow as the temperature 
increases is from bottom towards top. The vertical axis is $b$.}
\label{abflat}
\end{center}
\end{figure}

Figure (\ref{abflat}) shows the thermal path in the $(a-b)$ space. Note that there is only one path
for all values of $r_0$ as against Figure (\ref{ab}), where we have different paths for different values of $r_0$. This
is primarily due to the fact that here we have only one other scale, the cutoff $r_0$ or $R_c$ 
besides the temperature. So a change in $r_0$ will keep us on the same path as it can be 
compensated by a change in temperature. The fact that the transition temperature increases with $r_0$
is thus obvious as the phases depend on the dimensionless ratio $(\beta r_0/l^2)$.
This is not so for the theory discussed in the earlier part of this section where apart from $\beta$ 
we had both $R_c$ and $R_{S^3}$. In the numerical calculations here the critical value of $(a,b)$ is
$(0.681,0.664)$. It lies on the deconfinement line (dashed line in Figure (\ref{abflat})). 
For $r_0=0$ this value is never reached 
and the thermal path of $a-b$ always remains
above the deconfinement line. For rest of the finite values of $r_0$ the path crosses the deconfinement 
line and at low enough temperatures the values of $a$ and $b$ correspond to those in the confined phase.
However, since the unstable maximum is always below $0.5$, the curve for all values of $r_0$ never
crosses the third-order (black hole-string) transition line (solid line in Figure (\ref{abflat})) 
from above. The thermal history for nonzero temperature thus starts beyond this line.

\section{Discussion}\label{disc}

The analysis in this paper rests on the assumption that near the critical temperature, the
effective theory governing the phase transitions for ${\cal N}=4$ SYM on $S^3$ and the same
on $R^3$ with an IR cutoff are identical. Some evidence for this assumption is obtained from the phase 
transitions in gravity. On the gravitational side there is no phase transition for a theory
with boundary, $R^3$. The theory in this case is always in one of the phases; specifically 
the black hole free energy is less than that of thermal AdS for all non zero temperatures. 
However there are two ways to get a phase transition at finite temperature; one by changing
the boundary to $S^3$ and the other by putting a radial cutoff in gravity $r_0$, that has been 
discussed in section (\ref{grcutoff}). These result in lifting the free energy of the black hole 
above zero below a certain critical temperature. To this effect, the physical roles played by 
the cutoff $r_0$ and the finite radius of the three-sphere are identical. We have thus proposed
that the effective matrix model (\ref{effectiveaction}) should also model the phase transitions
for the theory with the cutoff. Comparing with the thermodynamic behaviour of its gravity dual,
we analyzed this matrix model. This model is parametrized by two parameters $a$ and $b$. We have
numerically calculated the variations of these parameters with respect to the temperature and the
cutoff. The relative behaviors of  $a$ and $b$ with respect to the scales reflect those
of the free energy of gravity. While for the theory on $S^3$ with a cutoff, due to the presence of 
three scales ($\beta$,$R_{S^3}$ and $R_c$), the nature of the relative dependences is not 
that clear, it
is quite simple when we have only two scales ($\beta$ and $R_c$).

Apart from the similarities of the two configurations, there are of course certain differences 
that we find in the analysing the response of the matrix model parameters to changes of temperature 
when considering  $S^3$ or $R^3$ with cutoff.
For the former, the parameters are increasing functions of temperature (see \cite{dmms1}). This
is not so in the latter case (Figure (\ref{atbtflat})). Also, from the computation of the potential
for a quark-antiquark pair, we find that the potentials differ in the above two cases. 
These differences could be due to the details of the form of the free energy from gravity. A weak coupling 
computation with an IR cutoff on the gauge theory side may shed some light on the 
similarities and differences of the two configurations.

\bigskip

\bigskip

\noindent{\bf Acknowledgements:}
We would like to thank Bala Sathiapalan for discussions and for his comments on the paper.
We also thank Souvik Banerjee for collaboration on results presented in the appendix.
The work of TKD is supported by the South African Research
Chairs Initiative of the Department of Science and Technology and National Research
Foundation. Any opinion, findings and conclusions or recommendations expressed in this
material are those of the authors and therefore the NRF and DST do not accept any
liability with regard thereto.

\appendix
\section{Dynamical cutoff?}

In (\ref{effstrongplane}), we 
saw that it was  
impossible to write down  a smooth potential which would interpolate 
between two minima (representing thermal AdS and black hole in the presence of cutoff) without 
having an unstable maximum. This maximum, in the gravity side, would represent a small
unstable black hole which is not seen in the hard wall model of \cite{herzog}. 
Furthermore, on generic ground, one may expect that since in this model, there is
a transition from AdS to black hole, there should be a bounce solution. The Lorentzian 
version of this bounce would be an unstable black hole. Can the absence of such a 
solution in thiese models be an artifact of introducing cutoff by hand? 
Within the framework of soft wall model, we try to address this issue in the appendix.
\footnote{This computations in this appendix was done along with Souvik Banerjee, Institute
of Physics, Bhubaneswar, India.}

Here, we consider a toy model in five dimensions which has a dilaton along with a 
non-trivial dilaton potential. This potential goes to 
a negative cosmological constant for small value of the scalar. We construct a family
of solutions for the metric with a soft wall like behaviour of the dilaton. At large
distance, the metric asymptotes to a flat AdS black hole. On the other hand, at short 
distance, due to dilaton backreaction, the metric gets deformed from it asymptotic structure. 
We indeed see that this 
deformation induces a small unstable black hole in the geometry. This perhaps suggests that had we 
dynamically intoduced a cutoff in the soft wall model of  \cite{herzog}, a small black 
hole would prehaps have showed up. However, it should be noted that, 
we only consider a toy model and it requires a fine-tuned dilaton potential. 

The model we consider is represented by the following action
\begin{equation}
S = \int d^5x{\sqrt{-g}}\Big(\frac{1}{2} R - \frac{1}{2} (\partial \phi)^2 - V(\phi)\Big).
\label{ac}
\end{equation}
We look for classical solution of this action where the metric has the
following form
\begin{equation}
dS^2 = e^{2 A} \Big( - h {dt^2} + \sum_{i = 1}^3 dx_i^2\Big) + \frac{e^{2 B}}{h} dr^2,
\label{metri}
\end{equation}
where $A, B, h$ are functions of $r$ only. The equations of motion 
can be written down by varrying the action with respect to the metric and the dilaton.
They are
\begin{eqnarray}
&&h^{\prime \prime} + ( 4 A^{\prime} - B^\prime) h^\prime = 0,\nonumber\\
&&A^{\prime \prime} - A^\prime B^\prime + \frac{1}{3} (\phi^\prime)^2 = 0,\nonumber\\
&&6 A^\prime h^\prime + h ( 24 A^{\prime ~2} - 2 \phi^{\prime ~2}) + 4 V e^{2 B} = 
0\nonumber\\
&&\phi^{\prime \prime} + \frac{h^\prime}{h} \phi^\prime + (4 A^\prime - B^\prime)\phi^\prime- 
\frac{e^{2B}}{ h} \frac{\partial V}{\partial \phi} = 0.
\label{eom}
\end{eqnarray}
Here the first three equations follow from the variation of the metric and the last
one is the scalar equation. Differentiating the third equation in (\ref{eom}) and using
the first two, we get the last equation. In this sense, there are only three
independent equations. Our will look for solutions of the above equations where the scalar 
depends on $r$ as 
\begin{equation}
\phi = \frac{a^2}{r^2}.
\label{dilaton}
\end{equation}
and then determine $A, B, h$ and $V$ by solving (\ref{eom}). 
It is easy to see that the following metric satisies the equations of motion
\begin{equation}
dS^2 = - r^2 \Big( h_0 + h_1 e^{-\frac{a^4}{3r^4}}\Big) dt^2 
+ \frac {e^{-\frac{2 a^4}{3r^4}}dr^2}{r^2 \Big( h_0 + h_1 
e^{-\frac{a^4}{3r^4}}\Big)} + r^2\sum_{i=1}^3 dx_i^2.
\label{sol}
\end{equation}
In the expression of the metric $h_0$ and $h_1$ are constants. The role of these
constants will be discussed shortly. The potential $V$ for which (\ref{sol}) is
a solution is given by
\begin{equation}
V = -2\Big[h_1\phi^2 e^{\frac{\phi^2}{3}} + (3 - \phi^2)( h_0 e^{\frac{2 \phi^2}{3}}
+ h_1 e^{\phi^2\over 3} )\Big].
\label{poten}
\end{equation}
We first note that for large $r$, as $\phi$ goes to zero, the potential goes to a negative 
constant given by
\begin{equation} 
V|_{r \rightarrow \infty} = - 6 (h_0 + h_1).
\end{equation}
Setting 
\begin{equation}
h_0 + h_1 = 1/l^2,
\label{rel1}
\end{equation}
we get, asymptotically,  the standard negative cosmological 
constant term ($-12/l^2$) in the action (\ref{ac}).  Furthermore, coefficient of $\phi^2$ in the 
potential, for $\phi \rightarrow 0$,  saturates the BF bound $ m = - 4/l^2$.
To leading order, the metric reduces to
\begin{equation}
dS^2 = -\Big(\frac{r^2}{l^2} - \frac{h_1 a^4}{3 r^2}\Big)dt^2 + 
\frac{dr^2}{\Big(\frac{r^2}{l^2} 
- \frac{h_1 a^4}{3 r^2}\Big)} + r^2\sum_{i=0}^3 dx_i^2,
\end{equation}
where we have used relation (\ref{rel1}). This is an AdS black hole solution with
flat horizon if we identify the mass of the black hole as 
\begin{equation}
M = \frac{h_1 a^4}{3}.
\end{equation}
Calculating $R_{\mu\nu\rho\sigma}R^{\mu\nu\rho\sigma}$, it is easy to check that (\ref{sol}) 
has a curvature singularity at $r = 0$.

The horizon of the metric is given by the zero of $h$. We denote the horizon radius as $r_+$.
It satisfies
\begin{equation}
r_+ = a \Big[\frac{1}{3 {\rm log}\big( {h_1l^2 \over{h_1l^2 -1}}\big)}\Big]^{1\over 4}.
\end{equation}

\begin{figure}[t]
\begin{center}
\begin{psfrags}
\psfrag{a}[][]{$r_+$}
\psfrag{b}[][]{$T(r_+)$}
\epsfig{file=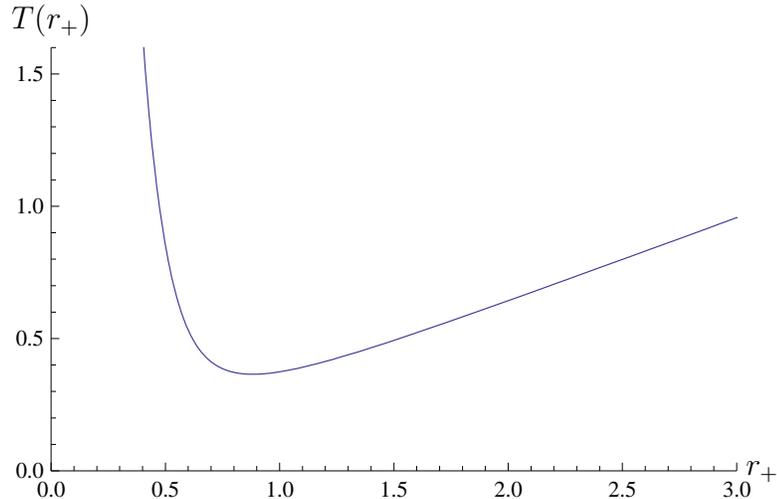, width= 10cm,angle=0}
\end{psfrags}
\vspace{ .1 in }
\caption{In this figure, we have shown the behaviour of $T$ as a function of $r_+$ for $l 
=1$, $a = 1$. 
Similar to the spherical AdS-Schwarzschild black hole, it shows that below a 
critical temperature, horizon does not exist.}
\label{tr}
\end{center}
\end{figure}

The temperature can be computed by analysing the  behaviour of the Euclidean version  
of the metric (\ref{sol}) near the horizon. For that, we 
define a new coordinate $\rho$ as 
\begin{equation}
h_0 + h_1 e^{-{a^4\over{3 r^4}}} = \rho^2.
\end{equation}
In terms of $\rho$, (\ref{sol}) takes the form
\begin{equation}
dS^2 = \frac{9 r_+^8}{4 h_1^2 a^8}\Big[ d\rho^2 + \frac{4 h_1^2 a^8 \rho^2}{9 r_+^6} 
d\tau^2\Big] + r_+^2 \sum_{i=1}^3 dx_i^2,
\end{equation}
where $\tau = - i t$. This metric does not have a conical singularity at $\rho = 0$ 
only when $\tau$ has a periodicity
\begin{equation}
\beta = \frac{3 \pi r_+^3}{h_1 a^4} = {3 \pi l^2\over {a^4}} \Big(1 - e^{-\frac {a^4}{3 
r_+^4}}\Big) r_+^3.
\end{equation}
Inverse of $\beta$ gives the temperature
\begin{equation}
T = \frac{a^4}{3 \pi l^2 \Big(1 - e^{-\frac {a^4}{3
r_+^4}}\Big) r_+^3}.
\label{temp}
\end{equation}
In the limit of large $r_+$, 
\begin{equation}
T \rightarrow \frac{r_+}{\pi l^2}.
\end{equation}
This is the temperature of a flat black hole in the absence of a scalar. 
For generic value of $r_+$, the behaviour of the temperature is shown
in figure (\ref{tr}).
We first not that the behaviour is reminiscent
to that of an AdS-Schearzschild black hole with sperical horizon. The temperature 
has a minimum when $r_+$ satisfies 
\begin{equation}
9 ( e^{a^4\over{3 r_+^4}} - 1) r_+^4 - 4 a^4 = 0.
\end{equation}
If we denote this value of $r_+$ as $r_{m}$, we find that the minimum temperature is
\begin{equation}
T_{m} = \frac{3 r_m e^{\frac{a^4}{3 r_m^4}}}{4 \pi l^2}.
\end{equation}
Temperature $T_m$ is identified as the nucleation temperature where a pair
of black hole nucleates. At any temperature above $T_m$, there are two black holes.
The smaller one is clearly unstable with negative specific heat -- it's temperature decreases 
as it increases in size. The bigger one, on the other hand, is always stable.

\end{document}